\documentclass[english,superscriptaddress,amssymb,twocolumn]{revtex4}
\newcommand{\modif}[1]{#1}

\usepackage{bm}
\usepackage{graphicx}
\usepackage{color}
\usepackage{amsmath}

\usepackage[T1]{fontenc}
\usepackage[latin9]{inputenc}
\usepackage{textcomp}
\usepackage{amssymb}
\usepackage{ulem}
\usepackage{extraplaceins}
\makeatother

\usepackage[english]{babel}

\begin{document}

\title{Persistent control of a superconducting qubit by stroboscopic measurement feedback}

\author{P. Campagne-Ibarcq}

\affiliation{Laboratoire Pierre Aigrain, Ecole Normale Sup\'erieure, CNRS (UMR 8551),
Universit\'e P. et M. Curie, Universit\'e D. Diderot 24, rue Lhomond,
75231 Paris Cedex 05, France }

\author{E. Flurin}

\affiliation{Laboratoire Pierre Aigrain, Ecole Normale Sup\'erieure, CNRS (UMR 8551),
Universit\'e P. et M. Curie, Universit\'e D. Diderot 24, rue Lhomond,
75231 Paris Cedex 05, France }

\author{N. Roch}

\affiliation{Laboratoire Pierre Aigrain, Ecole Normale Sup\'erieure, CNRS (UMR 8551),
Universit\'e P. et M. Curie, Universit\'e D. Diderot 24, rue Lhomond,
75231 Paris Cedex 05, France }

\author{D. Darson}

\affiliation{Laboratoire Pierre Aigrain, Ecole Normale Sup\'erieure, CNRS (UMR 8551),
Universit\'e P. et M. Curie, Universit\'e D. Diderot 24, rue Lhomond,
75231 Paris Cedex 05, France }

\author{P. Morfin}

\affiliation{Laboratoire Pierre Aigrain, Ecole Normale Sup\'erieure, CNRS (UMR 8551),
Universit\'e P. et M. Curie, Universit\'e D. Diderot 24, rue Lhomond,
75231 Paris Cedex 05, France }

\author{M. Mirrahimi}

\affiliation{INRIA Paris-Rocquencourt, Domaine de Voluceau, B.P. 105, 78153 Le Chesnay Cedex, France}

\author{M. H. Devoret}

\affiliation{Coll\`ege de France, 11 Place Marcelin Berthelot, F-75231 Paris Cedex
05, France }

\affiliation{Department of Applied Physics, Yale University, PO Box 208284, New
Haven, CT 06520-8284 }

\author{F. Mallet}

\affiliation{Laboratoire Pierre Aigrain, Ecole Normale Sup\'erieure, CNRS (UMR 8551),
Universit\'e P. et M. Curie, Universit\'e D. Diderot 24, rue Lhomond,
75231 Paris Cedex 05, France }

\author{B. Huard}

\affiliation{Laboratoire Pierre Aigrain, Ecole Normale Sup\'erieure, CNRS (UMR 8551),
Universit\'e P. et M. Curie, Universit\'e D. Diderot 24, rue Lhomond,
75231 Paris Cedex 05, France }

\date{\today}
\begin{abstract}
Making a system state follow a prescribed trajectory despite fluctuations and errors commonly consists in monitoring an observable (temperature, blood-glucose level \ldots) and reacting on its controllers (heater power, insulin amount \ldots). In the quantum domain, there is a change of paradigm in feedback since measurements modify the state of the system, most dramatically when the trajectory goes through superpositions of measurement eigenstates. Here, we demonstrate the stabilization of an arbitrary trajectory of a superconducting qubit by measurement based feedback. The protocol benefits from the long coherence time ($T_2>10~\mu$s) of the 3D transmon qubit, the high efficiency (82\%) of the phase preserving Josephson amplifier, and fast electronics ensuring less than 500~ns total delay. At discrete time intervals, the state of the qubit is measured and corrected in case an error is detected. For Rabi oscillations, where the discrete measurements occur when the qubit is supposed to be in the measurement pointer states, we demonstrate an average fidelity of 85\% to the targeted trajectory. For Ramsey oscillations, which does not go through pointer states, the average fidelity reaches $76\%$. Incidentally, we demonstrate a fast reset protocol allowing to cool a 3D transmon qubit down to $0.6\%$ in the excited state.
\end{abstract}

\maketitle

\section{Introduction}
The coupling of a quantum object to an environment is essential to enable its observation and manipulation. Yet, the mere existence of this coupling induces decoherence towards pointer states stable under monitoring of the environment~\cite{Zurek:2003p393}. There is thus a limiting timescale for the faithful preparation of a qubit in an arbitrary state, or its control along a given trajectory in Hilbert space. As a part of the environment, an observer extracts information on the object and contributes to this timescale. However, if the observer acquires information faster than the uncontrolled part of the environment, it is possible to use it through a feedback process and stabilize permanently a given trajectory or state~\cite{Doherty:2000p91, Wiseman2009, Sayrin2011,ZhouPRL2012}. Superconducting qubits in cavities offer a test bed for these concepts, as well as good candidates for practical applications~\cite{Devoret:2013p712,Clarke:2008p711}. Recently, persistent Rabi oscillations have been demonstrated via analog measurement based feedback using continuous weak measurement of a qubit~\cite{Vijay:2012fk}, and qubit reset via digital measurement based feedback using projective measurements has been performed~\cite{Riste:2012p707}. In this work, we demonstrate a simple protocol to stabilize any trajectory of a single qubit using a stroboscopic digital feedback based on strong measurement ~\cite{MayzarIEEE2012}. During the manipulation of the qubit, its state is measured in a nearly projective manner at specific time intervals and a correcting control sequence is triggered conditionally on the outcome so as to correct its trajectory from the errors due to decoherence and relaxation. The efficiency of the trajectory stabilization relies on the rapidity to measure and react compared to decoherence. In order to minimize these timescales, we used a phase preserving quantum limited amplifier~\cite{Bergeal:2010p331,Bergeal:2010p299,NicoPRL2012} and a Field Programmable Gate Array (FPGA) adding a delay of only 360~ns when outputing a drive pulse conditioned on readout (see supplementary material \cite{SupMatPCI2012} and references \cite{BarendsAPL2011,CorcolesAPL2011,SchusterPRL2005,Ong:2011p8032,Gambetta:2006p8229} therein).

\section{Fast and non-demolition projective measurement}

The superconducting qubit follows the design of the ``3D transmon'' developed in Ref.~\cite{Paik:2011p29}. A single aluminum Josephson junction, connected to two antennas of $0.4~$mm by $1~$mm each, on a sapphire substrate, is embedded in an empty bulk aluminum cavity whose first coupled modes are at $\omega_c/2\pi=7.748~\mathrm{GHz}$ and 13~GHz when the qubit is in its ground state. External coupling rates to the first mode $\kappa_{in}/2\pi=0.34~$MHz and $\kappa_{out}/2\pi=1.49~$MHz are chosen of the same order of magnitude as the inverse feedback delay (500~ns) and internal losses are negligible on these scales. The cavity is anchored to a dilution fridge below $30~$mK~\cite{SupMatPCI2012}. Spectroscopic measurements give a qubit frequency $\omega_{eg}=\omega_c-\Delta=2\pi\times 3.576~$GHz differing from the next transition by an anharmonicity $(\omega_{eg}-\omega_{fe})/2\pi=198~$MHz. The relaxation time $T_1= 28~\mu$s corresponds to the Purcell limit \cite{Houck:2008p411}, and pure dephasing time is $T_\phi= 14.5~\mu$s (Fig.~\ref{figure3}a). 
\begin{center}
\begin{figure*}
\includegraphics[scale=1.5]{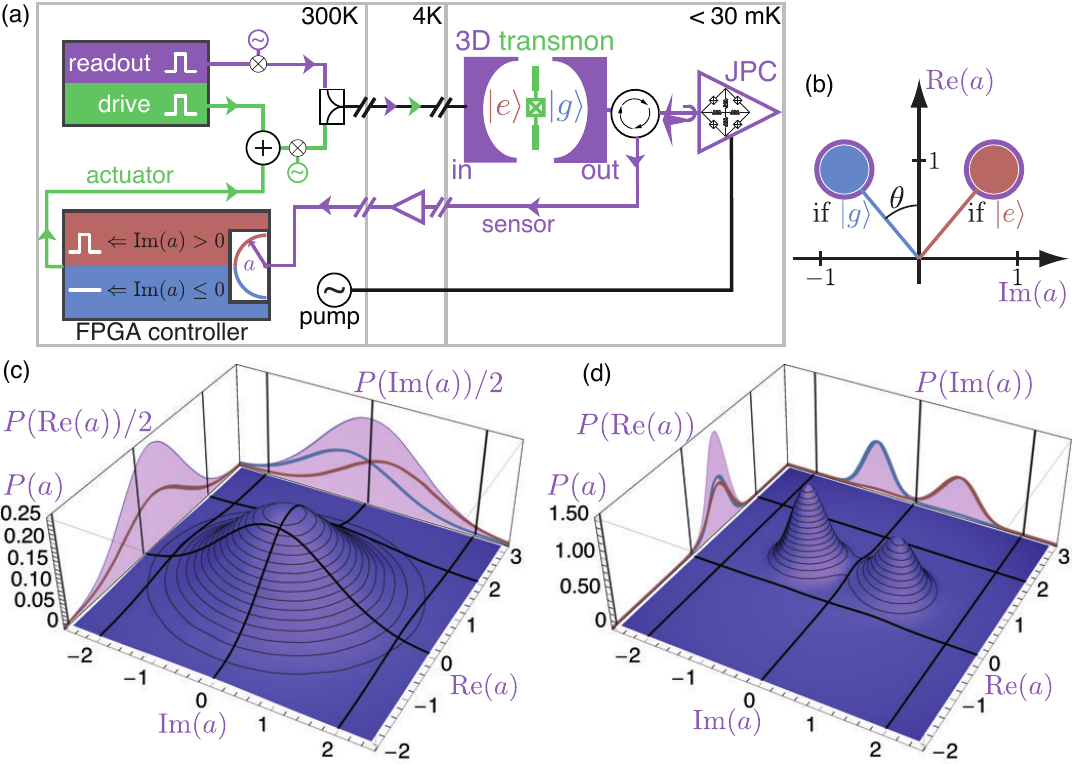}
\caption{ (\textbf{a}) Schematics of the experiment. The state of a 3D transmon qubit is entangled with the phase of a coherent field transmitted through the cavity at frequency $\omega_r=\omega_c-\chi$. It is amplified by a Josephson Parametric Converter (JPC) and its complex amplitude $a$ is measured and averaged by digital demodulation using an FPGA board (sensing and control). The drive at qubit transition frequency $\omega_{eg}$ is modulated by the sum of a predetermined waveform and of a conditional one generated by the FPGA board (actuation). If the transmission measurement points towards state $|e\rangle$, the actuator generates a $\pi$ pulse to get the qubit back in $|g\rangle$. (\textbf{b}) Expected complex amplitude of the field in the cavity averaged over the measurement time $T_\mathrm{meas}$ represented as the rod of a lollypop in the Fresnel space for both qubit states. The typical deviation due to vacuum fluctuations of the field in the $N_\mathrm{m}=11$ averaged modes is represented by the lollypop radius. The limited measurement efficiency ($\eta=67\%$) only slightly increases the observed deviations by $\eta^{-1/2}-1=22\%$ (purple rings). (\textbf{c,d}) Probability density with the JPC OFF (c) and ON (d), extracted from $10^6$ measurement outcomes when the qubit is prepared in states $|g\rangle$ or $|e\rangle$ with equal probability. Each outcome is the complex averaged amplitude of the field inside the transmon cavity at $\omega_r$. The halved probability density corresponding to the preparation of $|g\rangle$ only (resp. $|e\rangle$) is plotted in blue (orange) together with the projections along the real and imaginary axes. Turning on the pump of the amplifier (d) results in a great enhancement of the measurement fidelity compared to the case without (c).}
\label{figure1}
\end{figure*}
\end{center}

In the dispersive limit $\omega_{eg}-\omega_{fe}\ll\Delta$~\cite{Nigg:2012p710}, the cavity resonance frequency decreases by $2\chi$ when the qubit goes from ground to excited state, and the dispersive shift here is $\chi/2\pi=0.78~$MHz. The transmission measurement is strongest at readout frequency $\omega_r=\omega_c-\chi$ which minimizes the overlap between the two coherent states corresponding to the $|g\rangle$ and $|e\rangle$ qubit states (Fig.~\ref{figure1}). In the experiment, $1.2~\mu$s long square measurement pulses are sent through the cavity.  The amplitude of the readout field inside the cavity can be calibrated from the measurement induced dephasing as a function of readout power leading to 1.4 photons on average~\cite{SupMatPCI2012}. The outgoing signal is amplified during these $1.2~\mu$s using a Josephson Parametric Converter (JPC)~\cite{Bergeal:2010p331,Bergeal:2010p299,NicoPRL2012} with 22 dB of gain over 6~MHz (Fig.~\ref{figure1} and \cite{SupMatPCI2012}) and following amplifiers before being down-converted and digitalized using the FPGA board input. Note that the JPC was turned on only during measurement periods so as to minimize decoherence due to back-action~\cite{SupMatPCI2012}. The board averages numerically both quadratures of the signal during the steady part of the outgoing pulse only (see Fig.~3a), which corresponds to about $N_\mathrm{m}= T_\mathrm{meas}(\kappa_{in}+\kappa_{out})=11$ temporal modes of $1.4$ photons. States $|g\rangle$ and $|e\rangle$ for the qubit lead to two almost non overlapping coherent states for the average intracavity field $|\alpha e^{-i\theta}\rangle$ and  $|\alpha e^{i\theta}\rangle$ with $\theta\approx 40^\circ$ as expected from $\tan(\theta)=2\chi/(\kappa_{in}+\kappa_{out})$ (Fig.~\ref{figure1}b). With an ideal setup measuring both quadratures of the average complex field $a$ in the cavity, the variance on $a$ should be given by $1/\sqrt{N_\mathrm{m}}$~\cite{Caves:2012p310}. In the experiment, the $19\%$ loss of signal through the input of the cavity $(\kappa_{in}/\kappa_{tot})$ and the efficiency of the detection setup ($82\%$) degrade the signal by only $67\%$ beyond this variance (Fig.~\ref{figure1}). Therefore measuring $\mathrm{Im}(a)> 0$ on the readout field indicates a qubit in the excited state $|e\rangle$ with a fidelity beyond 99.8\%, taking aside the expected false counts due to relaxation events during readout. All measurement pulses in this manuscript are performed according to this procedure and the $0.2\%$ infidelity is neglected throughout. Using this setup, it is possible to perform almost projective and Quantum Non Demolition (QND) measurements of the qubit state much faster than decoherence \modif{\cite{Johnson:2012p705,Riste:2012p706}}, a crucial ingredient of measurement based feedback. An illustration of the discriminating power of the setup is shown as a histogram of measurement outcomes (average complex amplitude in the cavity) with the JPC amplifier on or off (Fig.~\ref{figure1}c,d) for a qubit starting randomly in state $|g\rangle$ or $|e\rangle$. 

\section{Cooling a qubit using measurement based feedback}

\begin{table}[htdp]
\caption{Error in the preparation of $|g\rangle$ using zero, one or two resets by feedback when starting in the most entropic state or in the thermalized state (effectively at $46~\mathrm{mK}$).}
\begin{center}
\begin{tabular}{|c|c|c|c|}
\hline
reset number & 0 & 1 & 2  \\
\hline
from $(|g\rangle\langle g|+|e\rangle\langle e|)/2$  & $50~\%$ & $3.6~\%$  & $1.1~\%$ \\
\hline
from thermalized state & $2.4~\%$  & $0.7~\%$  & $0.6~\%$ \\
\hline

\end{tabular}
\end{center}
\label{table1}
\end{table}

\begin{figure*}
\includegraphics[scale=.33]{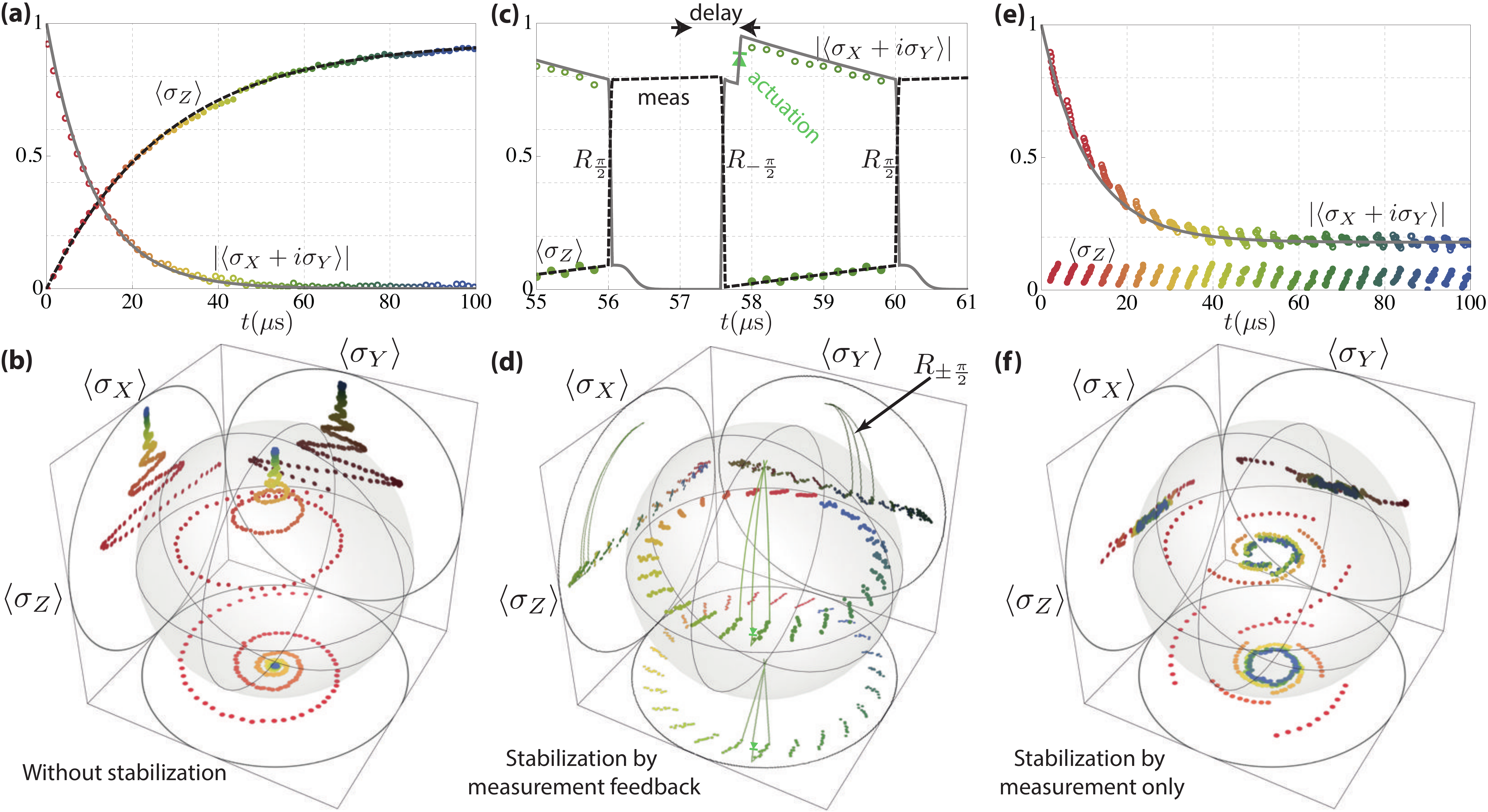}
\caption{(\textbf{a}) Evolution of $\langle\sigma_Z\rangle$ (dots) and of the coherence $|\langle\sigma_X+i\sigma_Y\rangle|$ (circles) when the qubit is prepared in state $(|g\rangle+|e\rangle)/\sqrt{2}$ at time $0$. The color encodes the time identically in all panels. Lines are exponential fits using coherence time $T_2= 11.5~\mu\mathrm{s}$ and  relaxation time $T_1=28~\mu\mathrm{s}$. (\textbf{b}) Same evolution represented in the Bloch sphere with a Ramsey frequency $\omega_{Ry}/2\pi=100~\mathrm{kHz}$. At each time (color), the outcome of qubit tomography is represented as a dot in the Bloch sphere and in the three orthogonal projection planes. The large black circles set the scale of the Bloch sphere extrema. (\textbf{c}) Same evolution as in (a) with stroboscopic measurement feedback every 4~$\mu$s. State tomography is only performed outside of the sensing and actuation periods. Lines represent results of a simulation without extra fit parameters. When the conditional $\pi$-pulse occurs (actuation), the average purity increases so that the coherences are preserved permanently on average. (\textbf{d}) Same evolution represented in the Bloch sphere with a Ramsey frequency $\omega_{Ry}/2\pi=10~\mathrm{kHz}$, instead of 100~kHz for a clearer observation of the trajectory. The simulated trajectory is represented as a line only for the time interval chosen in (c) for clarity.  (\textbf{e}) Evolution of the qubit with the same process as in (c,d) but without actuation. The exponential fit using the same $T_2$ as in (a) indicate an average persistent coherence of 18~\% without any actuation. (\textbf{f}) In the Bloch sphere, the Ramsey frequency is chosen to be $\omega_{Ry}/2\pi=100~\mathrm{kHz}$.}
\label{figure3}
\end{figure*}

As a \modif{benchmark of our feedback hardware, we actively} cool down the qubit \modif{to} its ground state, \modif{similarly to what was demonstrated by Rist\`e et al.} with a phase sensitive amplifier and digital controller~\modif{\cite{Riste:2012p707,Riste:2012Arxiv}}. Quantum information processing requires such removal of entropy during initialization or when correcting for errors \cite{DiVincenzo:2000p708}.  \modif{This} method allows to do so without fast frequency tuning \cite{Valenzuela:2006p702,Grajcar:2008p703,Reed:2010p143,Mariantoni:2011p704}, post-selection \cite{Johnson:2012p705,Riste:2012p706} or limited coupling rate $\kappa<\chi$ \cite{Geerlings:2012p709}. An initial measurement determines the qubit state. If the outcome points towards the excited state $(\mathrm{Im}(a)>0)$, the FPGA controller emits a square pulse (Fig.~\ref{figure1}) so as to apply a $\pi$-pulse around $Y$ on the qubit only $500~\mathrm{ns}$ after the first readout pulse exits the cavity (see \cite{SupMatPCI2012} for details). As an illustration, the qubit is first prepared in the most entropic mixed state $\rho=(|g\rangle\langle g|+|e\rangle\langle e|)/2$ by either applying a $\pi$-pulse or not (the outcomes are averaged over these two possibilities). The probability $P_{|e\rangle}$for the qubit to be in state $|e\rangle$ is then measured following zero, one or more resets by feedback. We found that starting from $P_{|e\rangle}=50~\%$, a single reset brings this level down to $P_{|e\rangle}=3.6~\%$ which would require to thermalize during $110~\mu\mathrm{s}$ without feedback. Yet events where the qubit relaxes between the middle of the measurement pulse and feedback pulse limit the efficiency of a single reset. Doing a second reset immediately after the first brings the qubit much closer to the ground state with $P_{|e\rangle}=1.1~\%$. This value does not improve with additional feedback and is limited mostly by the excitation of higher qubit states during the first reset~\modif{\cite{Riste:2012p707}}. These higher states are almost empty $(0.06~\%)$ when starting from a thermalized qubit at $P_{|e\rangle}=2.4~\%$ and two consecutive resets by feedback cool the qubit further down to $P_{|e\rangle}=0.6~\%$. These results are summarized in Table~\ref{table1}. Note that \modif{this reset allows} to prepare any state with similar purity using rotations of the qubit once the qubit is in state $|g\rangle$ \modif{and increase the repetition rate of quantum algorithms}\modif{~\cite{Riste:2012Arxiv}}.

\section{Stabilizing a quantum trajectory using stroboscopic feedback}

\subsection{Ramsey oscillations}

It is also possible to stabilize a state like $(|g\rangle+|e\rangle)/\sqrt{2}$ which is not an eigenstate of the measurement operator. First, a $\pi/2$-pulse is applied to the qubit so as to prepare it in  $(|g\rangle+ |e\rangle)/\sqrt{2}$ with a drive frequency $\omega_{eg}$. At any time $t$, it is possible to realize the full tomography of the qubit. Indeed, $\langle \sigma_Z\rangle$ is directly given by the average of the measurement outcomes, while $\langle \sigma_X\rangle$ (resp. $\langle \sigma_Y\rangle$) is given by the same averaging preceded by a rotation of the measurement axis using a $64$~ns long Rabi $\pi/2$-pulse around $Y$ (resp. $X$), where $\sigma_{X,Y,Z}$ are the Pauli matrices. In order to connect to the usual representation of Ramsey fringes at a given frequency $\omega_{Ry}$, we can rotate linearly in time the measurement axis so that $\langle \sigma_X\rangle$ maps onto $\langle \cos(\omega_{Ry}t) \sigma_X+\sin(\omega_{Ry}t)\sigma_Y\rangle$ and $\langle \sigma_Y\rangle$ onto $\langle -\sin(\omega_{Ry}t) \sigma_X+\cos(\omega_{Ry}t)\sigma_Y\rangle$. 

Without measurement based feedback, the Bloch vector of the qubit decays exponentially both in $Z$ and in the $X,Y$ plane (Fig.~\ref{figure3}a,b). The decay in $Z$ is described by timescale $T_1=28~\mu\mathrm{s}$ while the decay in $X,Y$ is described by timescale $T_2=11.5~\mu\mathrm{s}$. In order to stabilize persistent Ramsey oscillations, a measurement of  the qubit is performed after a $\pi/2$ rotation every 4~$\mu$s. The rotation axis is chosen so that the measurement outcome should point to state $|g\rangle$ in the targeted trajectory and the qubit is rotated back to its original state by a $-\pi/2$ pulse. Each time the qubit is found to be in the $|e\rangle$ state, the FPGA controller performs a fast $\pi$ pulse (actuation) with a delay of 500~ns after the measurement ends, which occurs after the $-\pi/2$ pulse. Using this stroboscopic measurement based feedback, Ramsey oscillations are indeed preserved indefinitely (Fig~\ref{figure3}c,d). Using optical Bloch  equations \cite{SupMatPCI2012}, one can calculate the predicted qubit trajectory corresponding to this protocol (Fig.~\ref{figure3}c) which is consistent with the experiment. Deviations to the experiment likely originate from the change in measurement induced dephasing when the JPC is turned off.  The average purity $\mathrm{Tr}(\rho^2)$ of the density matrix $\rho$ is calculated to be $85\%$ from these simulations, the time averaged fidelity $F=\overline{\langle\psi_\mathrm{targ}|\rho(t)|\psi_\mathrm{targ}\rangle}$ to the target trajectory $|\psi_\mathrm{targ}\rangle=(|g\rangle+e^{i\omega_{Ry}t} |e\rangle)/\sqrt{2}$ is $F=76\%$ and the average information quantity $1-\mathrm{Tr}(-\rho\mathrm{log}\rho)=0.60~\mathrm{bit}$. Interestingly, the sole effect of stroboscopically measuring the qubit, without any measurement feedback, induces persistent Ramsey oscillations, even though with less purity (52\%), fidelity (56\%) and information quantity ($0.03~\mathrm{bit}$) (Fig~\ref{figure3}e,f). This is due to the relaxation of the qubit during the measurement period towards state $|g\rangle$ making it more probable to reinitiate in state $(|g\rangle+|e\rangle)/\sqrt{2}$ than in state $(|g\rangle-|e\rangle)/\sqrt{2}$ after the measurement ends. This stabilization can be seen as a kind of reservoir engineering similar to \cite{Murch:2012p113} where the natural qubit decay is used as the dissipation source.

\subsection{Rabi oscillations}

\begin{figure}
\includegraphics[scale=.34]{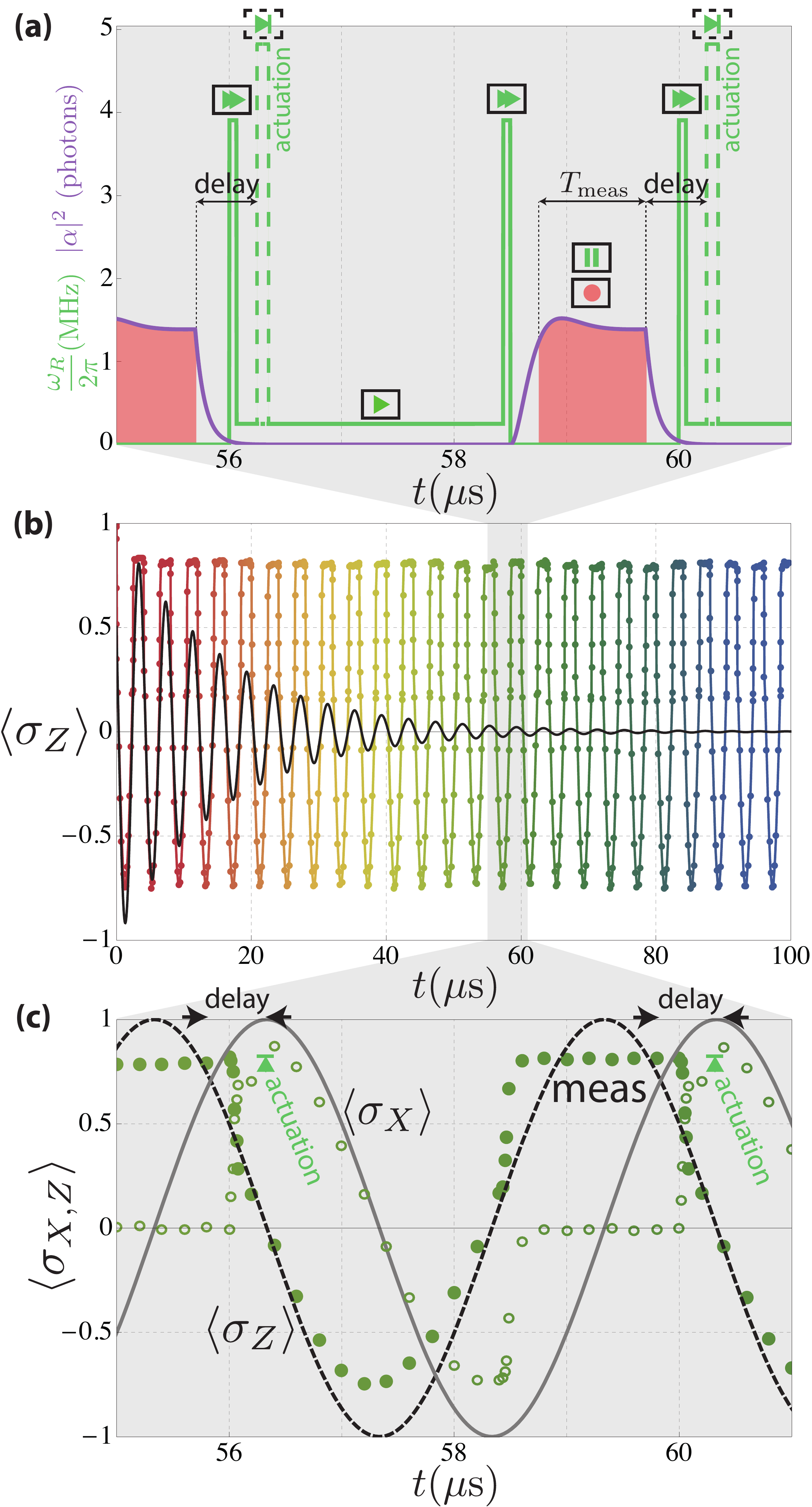}
\caption{(\textbf{a}) Pulse sequence for stabilizing Rabi oscillations. For a typical period of $4~\mu\mathrm{s}$, the lines represents the drive amplitude (green) and expected occupation of the cavity  (purple).  The complex amplitude $a$ of the measurement field is recorded only during the steady part of the occupation (red area). When $\mathrm{Im}(a)>0$, a fast $\pi$ pulse is applied after a total delay of $500~\mathrm{ns}$ (actuation). These steps are illustrated with usual symbols for media player. (\textbf{b}) Black line: decaying Rabi oscillation around $\sigma_Y$ with frequency $\omega_R=250~\mathrm{kHz}$ and measured decay time $T_R=15.5~\mu\mathrm{s}$. Dots on line: persistent Rabi oscillations measured using the pulse sequence described in (a). (\textbf{c}) Same measurement as in (b) shown on a smaller span for $\langle\sigma_Z\rangle$ (dots) and  $\langle\sigma_X\rangle$ (circles). The targeted Rabi trajectory is shown as two lines, dashed for $\langle\sigma_Z\rangle$ and gray for $\langle\sigma_X\rangle$.}
\label{figureRabi}
\end{figure}

In order to illustrate further the flexibility of stroboscopic projective measurement-based feedback, we have also stabilized Rabi oscillations. Although it is possible to perform this stabilization using analog feedback on a weak, continuous measurement~\cite{Vijay:2012fk}, we demonstrate here that discrete feedback events are more efficient~\cite{MayzarIEEE2012}. Without feedback, a constant microwave signal at $\omega_{eg}$ induces a Rabi oscillation of the qubit around $\sigma_Y$ with decay time $T_R= 15.5~\mu\mathrm{s}$ (Fig.~\ref{figureRabi}b) and frequency set to $\omega_R=250~\mathrm{kHz}$. In order to make the Rabi oscillations persistent, a measurement is performed each time the qubit is supposed to be in state $|g\rangle$ (Fig.~\ref{figureRabi}a). The FPGA controller then sends a fast correcting $\pi$-pulse (actuation) at the qubit frequency $\omega_{eg}$ each time the measurement reveals that the qubit is in the excited state. In order to optimize the fidelity of the feedback controlled trajectory to the targeted Rabi oscillation, the precession angle which is left idle during the measurement -- Zeno effect freezing the trajectory anyway -- is briefly accelerated before and after measurement to compensate exactly for that pause (see Fig.~\ref{figureRabi}a). As can be seen in Fig.~\ref{figureRabi}b, the Rabi oscillations are indeed stabilized permanently with this protocol. Their average fidelity to the targeted Rabi oscillation is $F=85\%$, their average purity 80\% and their average information quantity $0.50$~bit. The discrete correction events lead to visible discontinuities in the trajectories restoring the purity lost during the last Rabi period  due to decoherence.

\bigskip

\section{Conclusion}

The differences between continuous~\cite{Vijay:2012fk} and stroboscopic measurement feedback are enlightening. Though both methods allow the stabilization of a dynamical quantum state, continuous measurement exerts a constant dephasing rate while stroboscopic measurement allows variations of this rate in time. For trajectories like Rabi oscillations that go through eigenstates of the measurement observable (poles of the Bloch sphere), we benefit here from the versatility of stroboscopic feedback by measuring only close to state $|g\rangle$ which is insensitive to measurement induced dephasing, hence better preserving coherence over the whole trajectory. Besides, the stroboscopic method enables to stabilize trajectories like Ramsey oscillations, which never reach measurement eigenstates, by rotating periodically the measurement basis.

This work illustrates the possibilities offered by measurement-based feedback for circuit Quantum Electrodynamics in the case of a single qubit in cavity. We have shown here that fast digital electronics combined with efficient detection allow to realize elaborate quantum control protocols on these systems. Future error correction codes will benefit from the malleability of a numerical approach where complex filters need to be used to protect a quantum algorithm from errors. Extending these protocols to multi-qubit architectures should enable the preparation and stabilization of more complex entangled states and trajectories.

\begin{acknowledgments} We thank Hanhee Paik for discussions, Nathana\"el Cottet for help with electromagnetic field simulations, Jack Olejnik and Richard Pescari for realizing mechanical parts. Nanofabrication has been made within the consortium Salle Blanche Paris Centre and in the Quantronics group which we also thank. We thank the LERMA for slicing our Sapphire wafers. This work was supported by the EMERGENCES program "Contract" of Ville de Paris and by the ANR contract ANR-12-JCJC-TIQS.\end{acknowledgments}

\FloatBarrier


\begin{thebibliography}{10}

\bibitem{Zurek:2003p393}
W.~Zurek, ``Decoherence, einselection, and the quantum origins of the
  classical,'' {\em Rev. Mod. Phys.}, vol.~75, pp.~715--775, Jan 2003.

\bibitem{Doherty:2000p91}
A.~Doherty, S.~Habib, K.~Jacobs, H.~Mabuchi, and S.~Tan, ``Quantum feedback
  control and classical control theory,'' {\em Phys Rev A}, vol.~62, p.~012105,
  Jan 2000.

\bibitem{Wiseman2009}
H.~M. Wiseman and G.~J. Milburn, {\em Quantum Measurement and Control}.
\newblock Cambridge University Press, 2009.

\bibitem{Sayrin2011}
C.~Sayrin, I.~Dotsenko, X.~Zhou, B.~Peaudecerf, T.~Rybarczyk, S.~Gleyzes,
  P.~Rouchon, M.~Mirrahimi, H.~Amini, M.~Brune, J.-M. Raimond, and S.~Haroche,
  ``Real-time quantum feedback prepares and stabilizes photon number states,''
  {\em Nature}, vol.~477, pp.~73--77, Sep 2011.

\bibitem{ZhouPRL2012}
X.~Zhou, I.~Dotsenko, B.~Peaudecerf, T.~Rybarczyk, C.~Sayrin, S.~Gleyzes, J.~M.
  Raimond, M.~Brune, and S.~Haroche, ``Field locked to a fock state by quantum
  feedback with single photon corrections,'' {\em Phys. Rev. Lett.}, vol.~108,
  p.~243602, Jun 2012.

\bibitem{Devoret:2013p712}
M.~Devoret and J.~Martinis, ``Implementing qubits with superconducting
  integrated circuits,'' {\em Quantum Information Processing}, vol.~3, no.~1-5,
  pp.~163--203, 2004.

\bibitem{Clarke:2008p711}
J.~Clarke and F.~K. Wilhelm, ``Superconducting quantum bits,'' {\em Nature},
  vol.~453, p.~1031, Jun 2008.

\bibitem{Vijay:2012fk}
R.~Vijay, C.~Macklin, D.~H. Slichter, S.~J. Weber, K.~W. Murch, R.~Naik, A.~N.
  Korotkov, and I.~Siddiqi, ``Stabilizing rabi oscillations in a
  superconducting qubit using quantum feedback,'' {\em Nature}, vol.~490,
  no.~7418, pp.~77--80, 2012.

\bibitem{Riste:2012p707}
D.~Rist{\`e}, C.~Bultink, K.~Lehnert, and L.~Dicarlo, ``Feedback control of a
  solid-state qubit using high-fidelity projective measurement,'' {\em Phys.
  Rev. Lett.}, vol.~109, p.~240502, Dec 2012.

\bibitem{MayzarIEEE2012}
M.~Mirrahimi, B.~Huard, and M.~Devoret, ``Strong measurement and quantum
  feedback for persistent rabi oscillations in circuit qed experiments,'' {\em
 Proceedings of the 51st IEEE Conference on Decision and Control}, 2012.

\bibitem{Bergeal:2010p331}
N.~Bergeal, R.~Vijay, V.~E. Manucharyan, I.~Siddiqi, R.~J. Schoelkopf, S.~M.
  Girvin, and M.~H. Devoret, ``Analog information processing at the quantum
  limit with a josephson ring modulator,'' {\em Nat Phys}, vol.~6, p.~296, Feb
  2010.

\bibitem{Bergeal:2010p299}
N.~Bergeal, F.~Schackert, M.~Metcalfe, R.~Vijay, V.~E. Manucharyan, L.~Frunzio,
  D.~E. Prober, R.~J. Schoelkopf, S.~M. Girvin, and M.~H. Devoret,
  ``Phase-preserving amplification near the quantum limit with a josephson ring
  modulator,'' {\em Nature}, vol.~465, pp.~64--U70, Jan 2010.

\bibitem{NicoPRL2012}
N.~Roch, E.~Flurin, F.~Nguyen, P.~Morfin, P.~Campagne-Ibarcq, M.~H. Devoret,
  and B.~Huard, ``Widely tunable, nondegenerate three-wave mixing microwave
  device operating near the quantum limit,'' {\em Phys. Rev. Lett.}, vol.~108,
  p.~147701, Apr 2012.

\bibitem{Paik:2011p29}
H.~Paik, D.~Schuster, L.~Bishop, G.~Kirchmair, G.~Catelani, A.~Sears,
  B.~Johnson, M.~Reagor, L.~Frunzio, L.~Glazman, S.~Girvin, M.~Devoret, and
  R.~Schoelkopf, ``Observation of high coherence in josephson junction qubits
  measured in a three-dimensional circuit qed architecture,'' {\em Physical
  Review Letters}, vol.~107, p.~240501, Dec 2011.

\bibitem{SupMatPCI2012}
Supplementary Information attached.

\bibitem{Houck:2008p411}
A.~A. Houck, J.~A. Schreier, B.~R. Johnson, J.~M. Chow, J.~Koch, J.~M.
  Gambetta, D.~I. Schuster, L.~Frunzio, M.~H. Devoret, S.~M. Girvin, and R.~J.
  Schoelkopf, ``Controlling the spontaneous emission of a superconducting
  transmon qubit,'' {\em Physical Review Letters}, vol.~101, p.~080502, Jan
  2008.

\bibitem{Nigg:2012p710}
S.~Nigg, H.~Paik, B.~Vlastakis, G.~Kirchmair, S.~Shankar, L.~Frunzio,
  M.~Devoret, R.~Schoelkopf, and S.~Girvin, ``Black-box superconducting circuit
  quantization,'' {\em Physical Review Letters}, vol.~108, p.~240502, Jun 2012.

\bibitem{Caves:2012p310}
C.~M. Caves, J.~Combes, Z.~Jiang, and S.~Pandey, ``Quantum limits on
  phase-preserving linear amplifiers,''  {\em Physical Review A}, vol.~86, p.~063802, Dec 2012.
  
\bibitem{DiVincenzo:2000p708}
D.~DiVincenzo, ``The physical implementation of quantum computation,'' {\em
  Fortschr. Phys.}, vol.~48, no.~9-11, p.~771, 2000.

\bibitem{Valenzuela:2006p702}
S.~O. Valenzuela, W.~D. Oliver, D.~M. Berns, K.~K. Berggren, L.~S. Levitov, and
  T.~P. Orlando, ``Microwave-induced cooling of a superconducting qubit,'' {\em
  Science}, vol.~314, pp.~1589--1592, Dec 2006.

\bibitem{Grajcar:2008p703}
M.~Grajcar, S.~H.~W. van~der Ploeg, A.~Izmalkov, E.~Il'ichev, H.~G.
  Meyer, A.~Fedorov, A.~Shnirman, and G.~Sch\"on, ``Sisyphus cooling and
  amplification by a superconducting qubit,'' {\em Nature Physics}, vol.~4,
  p.~612, Jul 2008.

\bibitem{Reed:2010p143}
M.~D. Reed, B.~R. Johnson, A.~A. Houck, L.~Dicarlo, J.~M. Chow, D.~I. Schuster,
  L.~Frunzio, and R.~J. Schoelkopf, ``Fast reset and suppressing spontaneous
  emission of a superconducting qubit,'' {\em Applied Physics Letters},
  vol.~96, no.~20, p.~203110, 2010.

\bibitem{Mariantoni:2011p704}
M.~Mariantoni, H.~Wang, T.~Yamamoto, M.~Neeley, R.~C. Bialczak, Y.~Chen,
  M.~Lenander, E.~Lucero, A.~D. O'connell, D.~Sank, M.~Weides, J.~Wenner,
  Y.~Yin, J.~Zhao, A.~N. Korotkov, A.~N. Cleland, and J.~M. Martinis,
  ``Implementing the quantum von neumann architecture with superconducting
  circuits,'' {\em Science}, vol.~334, pp.~61--65, Oct 2011.

\bibitem{Johnson:2012p705}
J.~Johnson, C.~Macklin, D.~Slichter, R.~Vijay, E.~Weingarten, J.~Clarke, and
  I.~Siddiqi, ``Heralded state preparation in a superconducting qubit,'' {\em
  Phys. Rev. Lett.}, vol.~109, p.~050506, Aug 2012.

\bibitem{Riste:2012p706}
D.~Rist{\`e}, J.~V. Leeuwen, H.-S. Ku, K.~Lehnert, and L.~Dicarlo,
  ``Initialization by measurement of a superconducting quantum bit circuit,''
  {\em Phys. Rev. Lett.}, vol.~109, p.~050507, Aug 2012.

\bibitem{Riste:2012Arxiv}
D.~Rist{\`e}, C.~C. Bultink, M.~J. Tiggelman, R.~N. Schouten, K.~Lehnert, and L.~Dicarlo,
  ``Millisecond charge-parity fluctuations and induced decoherence in a superconducting qubit,''
  {\em arXiv.}, p.~1212.5459v1, Dec 2012.


\bibitem{Geerlings:2012p709}
K.~Geerlings, Z.~Leghtas, I.~M. Pop, S.~Shankar, L.~Frunzio, R.~J. Schoelkopf,
  M.~Mirrahimi, and M.~H. Devoret, ``Demonstrating a driven reset protocol of a
  superconducting qubit,'' {\em Phys. Rev. Lett.}, vol.~110, p.~120501, 2013

\bibitem{Murch:2012p113}
K.~W. Murch, U.~Vool, D.~Zhou, S.~J. Weber, S.~M. Girvin, and I.~Siddiqi,
  ``Cavity-assisted quantum bath engineering,'' {\em Phys. Rev. Lett.},
  vol.~109, p.~183602, 2012.

\expandafter\ifx\csname natexlab\endcsname\relax\def\natexlab#1{#1}\fi
\expandafter\ifx\csname bibnamefont\endcsname\relax
  \def\bibnamefont#1{#1}\fi
\expandafter\ifx\csname bibfnamefont\endcsname\relax
  \def\bibfnamefont#1{#1}\fi
\expandafter\ifx\csname citenamefont\endcsname\relax
  \def\citenamefont#1{#1}\fi
\expandafter\ifx\csname url\endcsname\relax
  \def\url#1{\texttt{#1}}\fi
\expandafter\ifx\csname urlprefix\endcsname\relax\def\urlprefix{URL }\fi
\providecommand{\bibinfo}[2]{#2}
\providecommand{\eprint}[2][]{\url{#2}}

\bibitem[{\citenamefont{Barends et~al.}(2011)\citenamefont{Barends, Wenner,
  Lenander, Y, Chen, Kelly, Lucero, O'Malley, Mariantoni, Sank
  et~al.}}]{BarendsAPL2011}
\bibinfo{author}{\bibfnamefont{R.}~\bibnamefont{Barends}},
  \bibinfo{author}{\bibfnamefont{J.}~\bibnamefont{Wenner}},
  \bibinfo{author}{\bibfnamefont{M.}~\bibnamefont{Lenander}},
  \bibinfo{author}{\bibnamefont{Y}},
  \bibinfo{author}{\bibfnamefont{B.}~\bibnamefont{Chen}},
  \bibinfo{author}{\bibfnamefont{J.}~\bibnamefont{Kelly}},
  \bibinfo{author}{\bibfnamefont{E.}~\bibnamefont{Lucero}},
  \bibinfo{author}{\bibfnamefont{P.}~\bibnamefont{O'Malley}},
  \bibinfo{author}{\bibfnamefont{M.}~\bibnamefont{Mariantoni}},
  \bibinfo{author}{\bibfnamefont{D.}~\bibnamefont{Sank}}, \bibnamefont{et~al.},
  \bibinfo{journal}{Appl. Phys. Lett.} \textbf{\bibinfo{volume}{99}},
  \bibinfo{pages}{113507} (\bibinfo{year}{2011}).
  
\bibitem[{\citenamefont{Corcoles et~al.}(2011)\citenamefont{Corcoles, Chow,
  Gambetta, Rigetti, Rozen, Keefe, Rothwell, Ketchen, and
  Steffen}}]{CorcolesAPL2011}
\bibinfo{author}{\bibfnamefont{A.~D.} \bibnamefont{Corcoles}},
  \bibinfo{author}{\bibfnamefont{J.~M.} \bibnamefont{Chow}},
  \bibinfo{author}{\bibfnamefont{J.~M.} \bibnamefont{Gambetta}},
  \bibinfo{author}{\bibfnamefont{C.}~\bibnamefont{Rigetti}},
  \bibinfo{author}{\bibfnamefont{J.~R.} \bibnamefont{Rozen}},
  \bibinfo{author}{\bibfnamefont{G.~A.} \bibnamefont{Keefe}},
  \bibinfo{author}{\bibfnamefont{M.~B.} \bibnamefont{Rothwell}},
  \bibinfo{author}{\bibfnamefont{M.~B.} \bibnamefont{Ketchen}},
  \bibnamefont{and} \bibinfo{author}{\bibfnamefont{M.}~\bibnamefont{Steffen}},
  \bibinfo{journal}{Appl. Phys. Lett.} \textbf{\bibinfo{volume}{99}},
  \bibinfo{pages}{181906} (\bibinfo{year}{2011}).

\bibitem[{\citenamefont{Schuster et~al.}(2005)\citenamefont{Schuster, Wallraff,
  Blais, Frunzio, Huang, Majer, Girvin, and Schoelkopf}}]{SchusterPRL2005}
\bibinfo{author}{\bibfnamefont{D.~I.} \bibnamefont{Schuster}},
  \bibinfo{author}{\bibfnamefont{A.}~\bibnamefont{Wallraff}},
  \bibinfo{author}{\bibfnamefont{A.}~\bibnamefont{Blais}},
  \bibinfo{author}{\bibfnamefont{L.}~\bibnamefont{Frunzio}},
  \bibinfo{author}{\bibfnamefont{R.-S.} \bibnamefont{Huang}},
  \bibinfo{author}{\bibfnamefont{J.}~\bibnamefont{Majer}},
  \bibinfo{author}{\bibfnamefont{S.~M.} \bibnamefont{Girvin}},
  \bibnamefont{and} \bibinfo{author}{\bibfnamefont{R.~J.}
  \bibnamefont{Schoelkopf}}, \bibinfo{journal}{Phys. Rev. Lett.}
  \textbf{\bibinfo{volume}{94}}, \bibinfo{pages}{123602}
  (\bibinfo{year}{2005}).

\bibitem[{\citenamefont{Ong et~al.}(2011)\citenamefont{Ong, Boissonneault,
  Mallet, Palacios-Laloy, Dewes, Doherty, Blais, Bertet, Vion, and
  Esteve}}]{Ong:2011p8032}
\bibinfo{author}{\bibfnamefont{F.~R.} \bibnamefont{Ong}},
  \bibinfo{author}{\bibfnamefont{M.}~\bibnamefont{Boissonneault}},
  \bibinfo{author}{\bibfnamefont{F.}~\bibnamefont{Mallet}},
  \bibinfo{author}{\bibfnamefont{A.}~\bibnamefont{Palacios-Laloy}},
  \bibinfo{author}{\bibfnamefont{A.}~\bibnamefont{Dewes}},
  \bibinfo{author}{\bibfnamefont{A.~C.} \bibnamefont{Doherty}},
  \bibinfo{author}{\bibfnamefont{A.}~\bibnamefont{Blais}},
  \bibinfo{author}{\bibfnamefont{P.}~\bibnamefont{Bertet}},
  \bibinfo{author}{\bibfnamefont{D.}~\bibnamefont{Vion}}, \bibnamefont{and}
  \bibinfo{author}{\bibfnamefont{D.}~\bibnamefont{Esteve}},
  \bibinfo{journal}{Phys. Rev. Lett.} \textbf{\bibinfo{volume}{106}},
  \bibinfo{pages}{167002} (\bibinfo{year}{2011}).

\bibitem[{\citenamefont{Gambetta et~al.}(2006)\citenamefont{Gambetta, Blais,
  Schuster, Wallraff, Frunzio, Majer, Devoret, Girvin, and
  Schoelkopf}}]{Gambetta:2006p8229}
\bibinfo{author}{\bibfnamefont{J.}~\bibnamefont{Gambetta}},
  \bibinfo{author}{\bibfnamefont{A.}~\bibnamefont{Blais}},
  \bibinfo{author}{\bibfnamefont{D.~I.} \bibnamefont{Schuster}},
  \bibinfo{author}{\bibfnamefont{A.}~\bibnamefont{Wallraff}},
  \bibinfo{author}{\bibfnamefont{L.}~\bibnamefont{Frunzio}},
  \bibinfo{author}{\bibfnamefont{J.}~\bibnamefont{Majer}},
  \bibinfo{author}{\bibfnamefont{M.~H.} \bibnamefont{Devoret}},
  \bibinfo{author}{\bibfnamefont{S.~M.} \bibnamefont{Girvin}},
  \bibnamefont{and} \bibinfo{author}{\bibfnamefont{R.~J.}
  \bibnamefont{Schoelkopf}}, \bibinfo{journal}{Phys. Rev. A}
  \textbf{\bibinfo{volume}{74}}, \bibinfo{pages}{042318}
  (\bibinfo{year}{2006}).

  
\end{thebibliography}

\end{document}